# Long-term yields of short-rate models as Feynman–Kac ground states: a variational quantum eigensolver with single-setting energy estimation on two NISQ platforms

Doyeol (David) Ahn

Department of Electrical and Computer Engineering, University of Seoul, Seoul 02504, Republic of Korea, and Singularity Quantum Inc., Newport Beach, CA 92660, USA. Correspondence: dahn@uos.ac.kr

## Abstract

We show that the long-term yield of a short-rate model is the ground-state energy of a Schrödinger operator. The operator is obtained from the Feynman–Kac bond-pricing equation by a Lamperti transform and a similarity transform. As a result, the long-term yield can be computed by a variational quantum eigensolver (VQE). We apply the method to the Vasicek, Cox–Ingersoll–Ross (CIR) and Black–Karasinski models. The last model has no closed-form solution. Calibrated monthly to the U.S. Treasury zero curve from 1985 to 2026, the resulting long-term yield follows the 30-year market yield with an in-sample correlation of 0.98. In noiseless simulation, a six-qubit real-amplitude ansatz reproduces the grid ground-state energy within 0.06 basis points (bp), and the grid error converges as $h^2$. The Feynman–Kac Hamiltonian is stoquastic. Therefore its ground state is positive, and the energy can be estimated from computational-basis samples alone, with a single measurement setting. On the IBM superconducting processor ibm_cleveland, we find that, for the binary encoding, gate noise appears mainly as bit flips of the high-order qubits, which move probability to the high-potential edges of the grid. We propose a tail-constrained bit-flip deconvolution in which the tail is fixed by a WKB criterion without using the exact solution. An ablation shows that this WKB tail set carries most of the correction. Combined with readout mitigation and zero-noise extrapolation, the four-qubit hardware estimates agree with the grid ground-state energies within 33–125 bp for all three models in ten runs over two sessions, and the distribution fidelity reaches 0.987–0.996. With parameters calibrated to the Treasury curve of October 2026, the errors with respect to the grid ground-state energies are 23–67 bp in a third session. On the IonQ trapped-ion processor Forte Enterprise, the same analysis reduces the error of the Vasicek circuit in the same proportions as on IBM, which shows that the bias structure is not specific to one platform. On the same circuits, a standard Pauli-measurement VQE with readout mitigation and zero-noise extrapolation deviates by 791–1,129 bp. Under these experimental conditions the single-setting estimator gives about ten times smaller errors, while it needs one measurement setting instead of nine. An on-device VQE trained with SPSA converges from a noiseless energy error of 4,939 bp to 130 bp in 150 iterations.

## 1. Introduction

Most proposals for quantum computing in finance address option pricing. Two approaches dominate. The first uses quantum amplitude estimation to accelerate Monte Carlo integration [1, 2]. The second maps the pricing partial differential equation onto a Schrödinger equation in imaginary time and evolves a variational state [3]. In this variational imaginary-time evolution, every time step requires the metric tensor and the energy gradient of the ansatz, which are measured with ancilla-controlled Hadamard tests [3, 26]. In both approaches the price is obtained from the time evolution of a specific initial state. On the other hand, the variational quantum eigensolver (VQE) [4] targets the ground state of a Hermitian operator. The ground state of the Black–Scholes operator carries no information on the option price. Therefore the VQE has found little use in derivative pricing.

In this work we identify a financial quantity that is a ground-state energy. For a short-rate model, the zero-coupon bond price $P(\tau, r)$ satisfies a Feynman–Kac equation with the short rate as a potential. After a Lamperti transform and a similarity transform, the generator becomes a Schrödinger operator $H$. For large maturity $\tau$ the bond price decays as $e^{-E_0\tau}$. Hence the long-term yield $y_\infty$ equals the ground-state energy $E_0$ of $H$. The variational imaginary-time evolution is a gradient flow that decreases the energy monotonically, and its limit is the ground state [26]. The long-term yield therefore needs only the endpoint of this flow, which the VQE targets directly, and none of the intermediate states. This spectral view of the term structure is well established in mathematical finance [5–7], and it is closely related to the path-integral formulation of interest-rate models in quantum finance [19, 20, 24]. It has not been used, to the author's knowledge, as a target for quantum ground-state algorithms.

The Feynman–Kac Hamiltonian has a second useful property. All its off-diagonal elements are non-positive, that is, it is stoquastic [8]. By the Perron–Frobenius theorem (Theorem 1 in Section 4.1) its ground state can be chosen positive, $\psi_0 = \sqrt{p_0}$. As a result, the energy can be estimated from computational-basis samples alone. Only one measurement setting is needed. This matches the time-based pricing of current superconducting processors, in which shots are inexpensive.

The contributions of this paper are as follows.

1. We formulate the long-term yield of one-factor short-rate models as the ground-state energy of a qubit Hamiltonian. We give the Lamperti and similarity transforms for the Vasicek, CIR and Black–Karasinski models and verify the CIR result against its closed form within $3 \times 10^{-5}$ bp.
2. We show that a shallow real-amplitude ansatz reproduces the grid ground state. In noiseless simulation the six-qubit error is 0.06 bp with 30 CNOT gates.
3. We introduce a single-setting energy estimator for stoquastic Hamiltonians. We diagnose that hardware gate noise appears mainly as bit flips of high-order qubits. We propose a tail constraint whose tail set follows from a WKB criterion without the exact solution, and we use it in a bit-flip deconvolution. Synthetic tests and an ablation show that the tail set carries most of the correction.
4. We demonstrate the method on the IBM processor ibm_cleveland for all three models. The four-qubit estimates agree with the grid ground-state energies within 33–125 bp in ten runs over two sessions. On the same circuits and under the conditions of Section 5.3, their errors are about ten times smaller than those of a standard Pauli-measurement VQE. Following Kandala et al. [16], we also train the ansatz on the device, with the $\sqrt{p}$ estimator as the cost function.
5. We calibrate the Vasicek and CIR models to the U.S. Treasury zero curve from 1985 to 2026. In sample, the long-term yield follows the 30-year market yield with a correlation of 0.98. With the parameters calibrated to the curve of October 2026, the hardware estimates agree with the grid ground-state energies within 23–67 bp, which demonstrates the complete chain from market data to a hardware estimate.

The paper is organized as follows. Section 2 derives the Feynman–Kac Hamiltonian and calibrates it to the U.S. Treasury curve. Section 3 describes the ansatz and the classical benchmarks. Section 4 develops the single-setting estimator and the error mitigation. Section 5 reports the hardware experiments. Sections 6–8 give the discussion, the limitations and the conclusion.

## 2. Feynman–Kac Hamiltonian and the long-term yield

### 2.1 Bond price as imaginary-time evolution

We assume a one-factor short-rate model under the risk-neutral measure,

$$dr_t = b(r_t)\,dt + s(r_t)\,dW_t. \tag{1}$$

Here $r_t$ is the short rate at time $t$, $b(r)$ is the drift, $s(r)$ is the diffusion coefficient and $W_t$ is a standard Brownian motion. The zero-coupon bond price with maturity $\tau$ is $P(\tau, r) = \mathbb{E}\big[\exp\big(-\int_0^\tau r_t\ dt\big)\,\big|\,r_0 = r\big]$. By the Feynman–Kac theorem, it satisfies

$$\frac{\partial P}{\partial \tau} = \mathcal{A}P, \qquad \mathcal{A} = \frac{1}{2}s^2(r)\frac{\partial^2}{\partial r^2} + b(r)\frac{\partial}{\partial r} - r, \qquad P(0, r) = 1. \tag{2}$$

Here $\tau$ is the time to maturity, $\mathcal{A}$ is the generator of the process (1) with the discount term $-r$, and the initial condition $P(0, r) = 1$ is the unit payoff at maturity.

### 2.2 Lamperti and similarity transforms

In order to obtain a constant diffusion coefficient $s_0$, we introduce the Lamperti coordinate $y = s_0 \int dr/s(r)$. By Itô's lemma, $y$ has the drift $\tilde{b}(y)$ and the diffusion $s_0$. We then write $P = e^{-W}\psi$ with $W(y) = \int \tilde{b}(y)/s_0^2\,dy$. By substituting this into Eq. (2), the first-derivative term cancels and we obtain $\partial\psi/\,\partial\tau = -H\psi$ with

$$H = -\frac{s_0^2}{2}\frac{\partial^2}{\partial y^2} + V(y), \qquad V(y) = r(y) + \frac{\tilde{b}(y)^2}{2s_0^2} + \frac{\tilde{b}'(y)}{2}. \tag{3}$$

Here $V(y)$ is the effective potential, $r(y)$ is the short rate expressed in the Lamperti coordinate, and $\tilde{b}'(y) = d\tilde{b}/dy$. The operator $H$ is Hermitian with respect to the flat measure $dy$. Expanding in its eigenstates $H\phi_k = E_k\phi_k$, we obtain

$$P(\tau, y) = e^{-W(y)} \sum_k e^{-E_k\tau}\ \phi_k(y)\,\langle\phi_k|e^W\rangle\ \to\ c_0\, e^{-W(y)}\phi_0(y)\, e^{-E_0\tau} \quad (\tau \to \infty). \tag{4}$$

Here $E_k$ and $\phi_k$ are the eigenvalues and the normalized eigenfunctions of $H$, with $E_0 < E_1 \le E_2 \le \cdots$, and $c_0 = \langle\phi_0|e^W\rangle$. Therefore the long-term yield $y_\infty = \lim_{\tau\to\infty}[-\ln P(\tau)/\tau]$ equals the ground-state energy $E_0$. The instantaneous forward rate $f(\tau) = -\,\partial \ln P/\,\partial\tau$ [26] converges to the same limit. By Eq. (4), $f(\tau) - E_0$ decays as $e^{-(E_1-E_0)\tau}$, whereas the yield approaches $E_0$ as $1/\tau$ because of the constant prefactor in Eq. (4). Hence the spectral gap $E_1 - E_0$ sets the rate at which the forward curve flattens, and $E_0$ fixes the long end of both the yield curve and the forward curve. This is the operator view of long-term risk of Hansen and Scheinkman [5] and the spectral expansion of Linetsky [6].

The same result follows from the path-integral form of the Feynman–Kac formula. Kac [18] showed that Feynman's path integral [17] becomes a well-defined Wiener integral in imaginary time. The bond price is then a sum over short-rate paths,

$$P(\tau) = \int \mathcal{D}y\ e^{-S_E[y]}, \qquad S_E[y] = \int_0^\tau \left[ \frac{\left(\dot{y} - \tilde{b}(y)\right)^2}{2s_0^2} + r(y) \right] dt. \tag{5}$$

Here $\mathcal{D}y$ denotes the measure over paths $y(t)$ that start at $y(0) = y$, $S_E$ is the Euclidean action and $\dot{y} = dy/dt$. By expanding the square, the term $\dot{y}^2/(2s_0^2)$ gives the kinetic operator and the term $\tilde{b}^2/(2s_0^2)$ enters the potential. The cross term $-\tilde{b}\dot{y}/s_0^2$ is a total derivative. It gives the boundary factors $e^{-W}$ of the similarity transform. The term $\tilde{b}'/2$ is the Itô correction that arises when the path integral is discretized with the pre-point rule [19, 21]. The derivation is given in Appendix A. Hence the short rate acts as a potential accumulated along each path. In probabilistic terms, it is the killing rate of the diffusion, and $E_0$ is the asymptotic decay rate of the discounted weight. The long-term yield plays the role of the mass that is extracted from the decay of a Euclidean correlation function in lattice field theory. Path integrals and Hamiltonians for interest-rate models have been developed extensively by Baaquie and co-workers in quantum finance [19, 20, 24]. In this work we use the Hamiltonian as the target of a quantum ground-state algorithm.

### 2.3 Models

**Vasicek** [9]. Here $b = \kappa(\theta - r)$ and $s = \sigma$, so $y = r$ and $s_0 = \sigma$. The potential

$$V(r) = r + \frac{\kappa^2(\theta - r)^2}{2\sigma^2} - \frac{\kappa}{2} \tag{6}$$

is a shifted harmonic oscillator with frequency $\kappa$. Here $\kappa$ is the speed of mean reversion, $\theta$ is the long-run mean of the short rate and $\sigma$ is the volatility. It can be shown that $E_0 = \theta - \sigma^2/(2\kappa^2)$, which is the known long-term Vasicek yield.

**Cox–Ingersoll–Ross** [10]. Here $b = \kappa(\theta - r)$ and $s = \sigma\sqrt{r}$. The Lamperti coordinate is $y = 2\sqrt{r}/\sigma$ with $s_0 = 1$ and $\tilde{b} = c/y - \kappa y/2$, where $c = (4\kappa\theta - \sigma^2)/(2\sigma^2)$. We obtain

$$V(y) = \frac{c(c-1)}{2y^2} + \left(\frac{\sigma^2}{4} + \frac{\kappa^2}{8}\right)y^2 - \frac{c\kappa}{2} - \frac{\kappa}{4}, \qquad E_0 = \frac{2\kappa\theta}{\kappa + \gamma}, \qquad \gamma = \sqrt{\kappa^2 + 2\sigma^2}. \tag{7}$$

Here $\kappa$ and $\theta$ have the same meaning as in the Vasicek model, $\sigma$ is the volatility coefficient of $\sqrt{r}$, and $\gamma$ is the parameter that appears in the closed-form CIR bond price. The first term is a centrifugal term. It is repulsive when $c > 1$, that is, when $4\kappa\theta > 3\sigma^2$. This condition is slightly stronger than the Feller condition $2\kappa\theta \geq \sigma^2$.

**Black–Karasinski** [11]. Here $d\ln r = \kappa(\theta_l - \ln r)\, dt + \sigma\, dW$. With $x = \ln r$ and $s_0 = \sigma$, the potential is

$$V(x) = e^x + \frac{\kappa^2(\theta_l - x)^2}{2\sigma^2} - \frac{\kappa}{2}. \tag{8}$$

Here $\theta_l$ is the long-run mean of $\ln r$, and $\kappa$ and $\sigma$ are the speed of mean reversion and the volatility of $\ln r$. The exponential term makes the problem non-affine, and no closed form for $E_0$ is known.

Throughout we use $\kappa = 0.3$ and a mean level of 5%. The volatilities are $\sigma = 2\%$ (Vasicek), $\sigma = 5\%$ (CIR, $c = 11.5$) and $\sigma = 25\%$ (Black–Karasinski, $\bar{r} = e^{\theta_l} = 5\%$).

### 2.4 Calibration to the U.S. Treasury curve

In order to relate $E_0$ to market data, we calibrated the Vasicek and CIR models to the zero-coupon U.S. Treasury curve of Gürkaynak, Sack and Wright [28], which the Federal Reserve Board publishes daily. For each month from November 1985 to October 2026 (492 months), we fitted the parameters $(r_0, \kappa, \theta, \sigma)$ to the continuously compounded zero yields for maturities of 1 to 30 years, with the mean-reversion speed bounded by $\kappa \geq 0.1$. The ground-state energy $E_0$ and the gap $E_1 - E_0$ then follow from the Hamiltonian of Eq. (3). The details and the robustness checks are given in Appendix E.

Figure 1 shows the result. The curves are fitted with a median error of 7.5 bp for Vasicek and 8.3 bp for CIR. Over 41 years, $E_0$ follows the 30-year zero yield with a correlation of 0.98, and it lies above it by about 50 bp. This correlation reflects the in-sample consistency of the model-implied long-term yield with the calibrated curves; it is not an out-of-sample forecast of future yields. The gap has a median of 0.23 per year for Vasicek and 0.25 per year for CIR. This corresponds to a time scale of about 4 years for the flattening of the forward curve. The calibrated long-term yields range from about 2% to 11%. Hence the parameters used in the rest of this work, $\kappa = 0.3$ and a mean level of 5%, lie within the range observed in the market. The value of $E_0$ depends on the admissible range of $\kappa$ and on the extrapolation of the model beyond 30 years, and the Black–Karasinski fit is unreliable at near-zero short rates (Appendix E). In Section 5.6 we also evaluate the four-qubit circuits with the parameters calibrated to the curve of 16 October 2026 on quantum hardware.

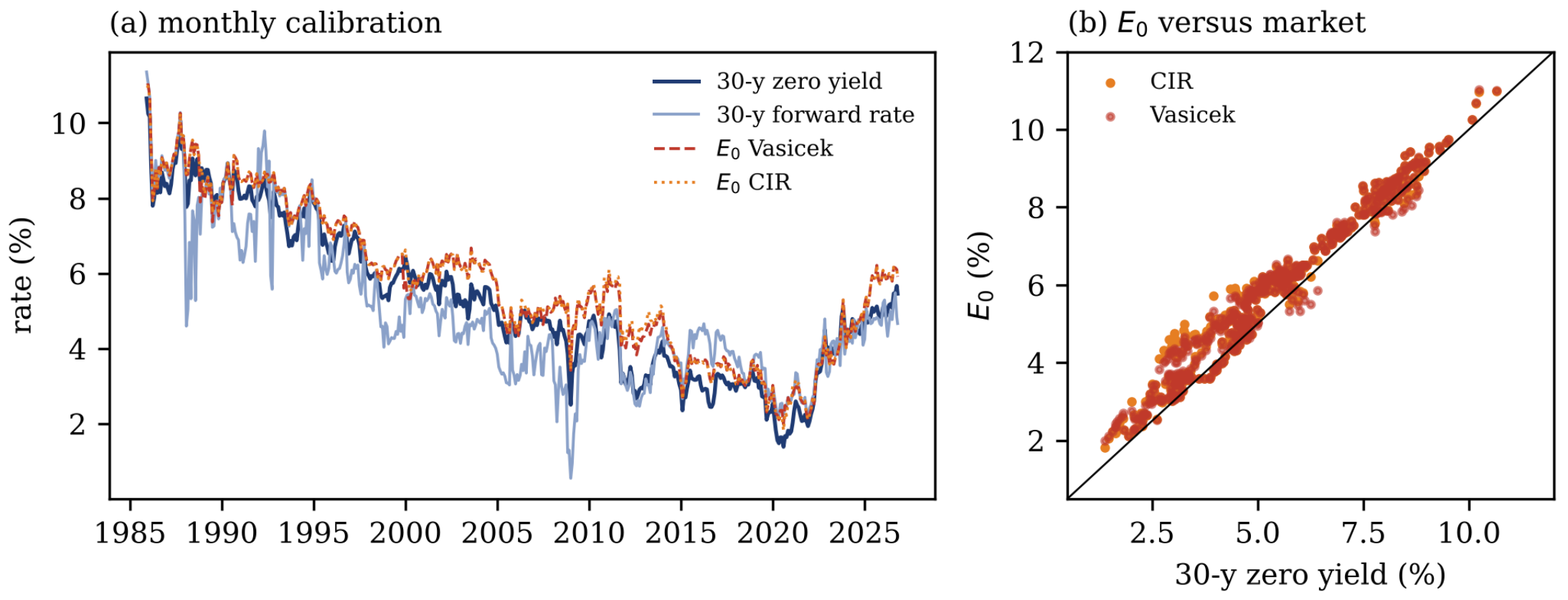

***Figure 1.*** *Long-term yield $E_0$ of the Vasicek and CIR models calibrated monthly to the U.S. Treasury zero curve [28], November 1985 to October 2026, with $\kappa \geq 0.1$. (a) $E_0$ together with the 30-year zero yield and the 30-year instantaneous forward rate. (b) $E_0$ versus the 30-year zero yield; the line is the diagonal.*

### 2.5 Discretization and qubit encoding

We discretize $y$ on $2^n$ equally spaced points with spacing $h$ and Dirichlet boundaries. The second-order finite-difference Hamiltonian is

$$H = \mathrm{diag}(d_0, \dots, d_{2^n-1}) + e\left(S + S^\top\right), \qquad d_j = \frac{s_0^2}{h^2} + V(y_j), \qquad e = -\frac{s_0^2}{2h^2} < 0, \tag{9}$$

where $y_j = y_0 + jh$ $(j = 0, \dots, 2^n - 1)$ are the grid points, $d_j$ and $e$ are the diagonal and off-diagonal elements, and $S$ is the lower shift matrix with $S_{j+1,j} = 1$. Since $d_j = 2|e| + V(y_j)$, the operator $-H$ is the generator of a continuous-time random walk on the grid, which hops to each neighbor at the rate $|e|$, combined with the potential $V$. Equation (9) is therefore the lattice counterpart of the continuum limit that leads from a random walk to the diffusion equation [25]. The grid index $j$ is encoded in binary on $n$ qubits, with qubit 0 as the most significant bit. The grid windows are $\pm 5\ell$ (Vasicek) and $\pm 6\ell$ (Black–Karasinski) about the potential minimum, with $\ell = \sigma/\sqrt{\kappa}$. For CIR the hardware grid is a window of $\pm 5$ standard deviations about the ground state, away from the $1/y^2$ wall; its truncation error is 0.1 bp.

The Pauli decomposition of $H$ at $n = 6$ contains 79 strings for Vasicek. Of these, 16 are diagonal (the identity and 15 $ZZ$ terms, since $V$ is quadratic in $y$) and 63 are off-diagonal. For CIR and Black–Karasinski the non-polynomial potentials give 64 diagonal strings. In general the kinetic term contributes $2^n - 1$ off-diagonal strings. This growth motivates the single-setting estimator of Section 4.

# 3. Variational algorithm and classical benchmarks

## 3.1 Ansatz and optimization

The ground state of $H$ is real and positive. Hence we use a real-amplitude ansatz. Each of $L$ layers applies $R_y(\theta)$ to every qubit and then a ladder of CNOT gates on the pairs $(q, q+1)$. A final $R_y$ layer follows. The ansatz has $n(L+1)$ parameters and $L(n-1)$ CNOT gates, and it maps directly onto a linear chain of physical qubits. The four-qubit circuit with $L = 3$ is shown in Appendix F. For the classical benchmarks we minimize the exact energy $\langle \psi(\boldsymbol{\theta}) | H | \psi(\boldsymbol{\theta}) \rangle$ with L-BFGS. The gradient is obtained by adjoint differentiation, and we take the best of several random initializations.

## 3.2 Results

Figure 2 summarizes the Vasicek benchmark at $n = 6$. The VQE state coincides with the continuum ground state (Fig. 2a). The energy error with respect to the grid ground state falls from 220 bp at $L = 3$ to 3.3 bp at $L = 4$, 0.058 bp at $L = 6$ (fidelity 0.99999963, 30 CNOT gates) and 0.0002 bp at $L = 8$ (Fig. 2b). At $n = 6$ the VQE error is already below the discretization error of the grid, $-2.4$ bp. The discretization error converges as $h^2$ and reaches $-0.14$ bp at $n = 8$ and $-0.009$ bp at $n = 10$ (Fig. 2c).

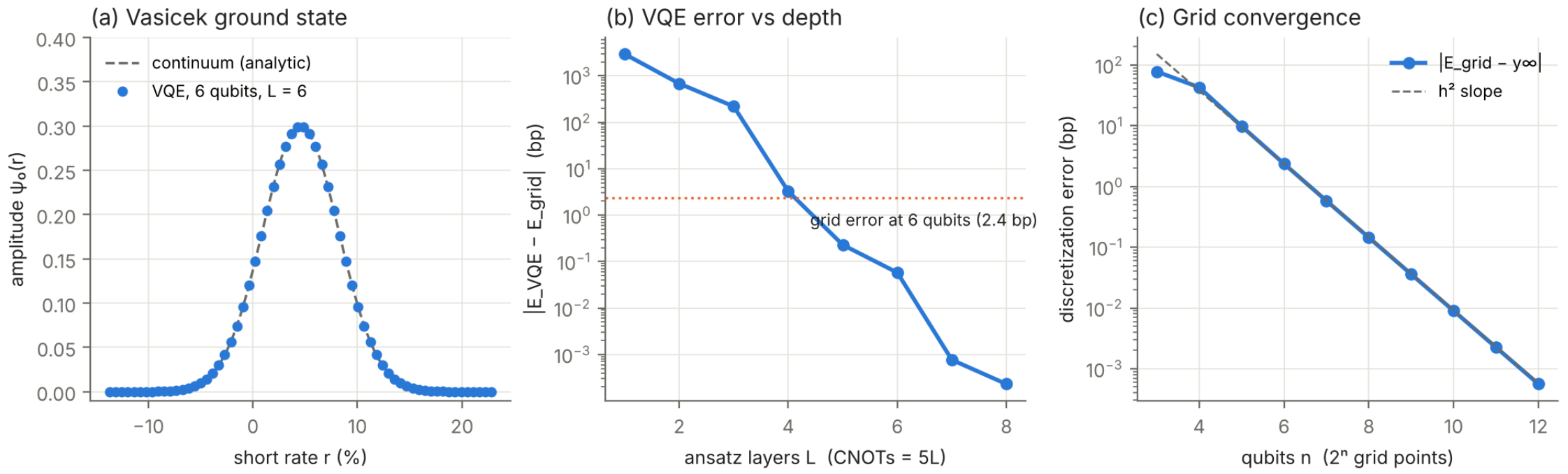

***Figure 2.*** *Vasicek benchmark ($\kappa = 0.3$, $\theta = 5\%$, $\sigma = 2\%$) in noiseless simulation. (a) Ground-state amplitude from the six-qubit VQE with $L = 6$ (dots) and the continuum Gaussian ground state (dashed). (b) VQE energy error with respect to the grid ground state versus the number of ansatz layers L; the dotted line marks the grid discretization error at six qubits. (c) Discretization error $|E_{\text{grid}} - y_\infty|$ versus the number of qubits, with an $h^2$ reference slope.*

Table 1 lists the results for the three models. The same code reproduces the closed-form CIR yield, which confirms the Lamperti and similarity transforms for a state-dependent diffusion. For Black–Karasinski the fine-grid value 5.17200% serves as the reference. At four qubits, three layers (9 CNOT gates) reproduce the grid ground state exactly for all three models. These four-qubit circuits are used on hardware in Section 5.

**Table 1.** Long-term yields in noiseless simulation. Errors are in basis points. The six-qubit CIR entries use a grid that starts at $y = 0$; the four-qubit CIR entries use the windowed hardware grid of Section 2.5.

| Model | Reference $y_\infty$ | $E_0$, fine grid ($2^{14}$ points) | 6 qubits: VQE ($L = 6$) − grid | 6 qubits: grid − reference | 4 qubits: grid $E_0$ | 4 qubits: grid − reference |
|---|---|---|---|---|---|---|
| Vasicek | 4.77778% (closed form) | 4.77778% | +0.058 | −2.37 | 4.34843% | −42.9 |
| CIR | 4.93242% (closed form) | 4.93242% | +0.92 | −2.37 | 4.70775% | −22.5 |
| Black–Karasinski | none | 5.17200% | +0.12 | −3.53 | 4.52259% | −64.9 |

# 4. Single-setting energy estimation for stoquastic Hamiltonians

## 4.1 The $\sqrt{p}$ estimator

The off-diagonal element $e$ of $H$ is negative, so that $H$ is stoquastic. Moreover, $H$ is irreducible, since every grid point is coupled to its neighbors. We then use the following theorem [27].

**Theorem 1 (Perron–Frobenius).** *Let $H$ be a real symmetric matrix whose off-diagonal elements are non-positive, and let $H$ be irreducible. Then the lowest eigenvalue $E_0$ of $H$ is nondegenerate, and its eigenvector can be chosen with strictly positive components.*

The proof is given in Appendix C. By Theorem 1, the ground state is $\psi_0 = \sqrt{p_0}$ with $p_{0,j} > 0$ for all $j$, and the gap $E_1 - E_0$ is strictly positive. For a measured distribution $p$ over the grid we define

$$E[p] = \langle\sqrt{p}\,|\,H\,|\,\sqrt{p}\rangle = \sum_j d_j\ p_j + 2e \sum_j \sqrt{p_j\, p_{j+1}}. \tag{10}$$

Here $p_j$ is the measured probability of grid point $j$, and the second sum runs over $j = 0, \dots, 2^n - 2$. This estimator needs only computational-basis samples, that is, one measurement setting. It is exact when $p = p_0$. Moreover, $\sqrt{p}$ is a normalized state for any distribution $p$. By the variational principle, $E[p] \geq E_0$ for every $p$, including a finite-shot histogram. Therefore noise and sampling errors can only raise the estimate. We note that $E[p]$ is the energy of the state $\sqrt{p}$ with non-negative amplitudes. It is not, in general, the expectation value of $H$ in the prepared state $|\psi\rangle$, whose amplitudes may have arbitrary signs. For a stoquastic $H$ and $p_j = |\psi_j|^2$, the first step of Appendix C gives $E_0 \leq E[p] \leq \langle\psi|H|\psi\rangle$. Hence the estimator is a variational bound on $E_0$ for any prepared state, and it equals the energy of the prepared state when its amplitudes are non-negative. For a Hamiltonian that is not stoquastic, Eq. (10) does not apply.

The estimator is sensitive to probability that leaks into the tails. If a small weight $\delta p_j$ appears where $p_0$ vanishes, the energy rises to first order by approximately $\delta p_j(d_j - E_0)$. Since $d_j$ grows with the potential, leakage into the grid edges is the most harmful.

### 4.2 Readout mitigation

We prepare $|0 \cdots 0\rangle$ and $|1 \cdots 1\rangle$ on the same physical qubits and estimate, for each qubit, the probabilities $P(1|0)$ and $P(0|1)$. For $n \leq 6$ the tensored $2^n \times 2^n$ confusion matrix is inverted exactly. The resulting quasi-probabilities are projected onto the probability simplex.

### 4.3 Tail-constrained bit-flip deconvolution

In the binary encoding a flip of qubit $q$ moves the grid index by $\pm 2^{n-1-q}$. The ground state is concentrated near the middle of the grid, at $j = 2^{n-1} - 1$ and $2^{n-1}$. A single flip of the most significant qubit therefore sends this weight to $j = 2^n - 1$ and $j = 0$, where $d_j$ is largest. We model the residual gate noise as independent effective bit flips,

$$p^{\text{meas}} = \left(\otimes_{q=0}^{n-1} B(\varepsilon_q)\right) p, \qquad B(\varepsilon) = \begin{pmatrix} 1-\varepsilon & \varepsilon \\ \varepsilon & 1-\varepsilon \end{pmatrix}. \tag{11}$$

Here $p$ is the distribution before the effective bit flips, $p^{\text{meas}}$ is the measured distribution, both in bit order, and $B(\varepsilon_q)$ is the flip matrix of qubit $q$ with flip probability $\varepsilon_q$. The rates $\varepsilon_q$ are not known. We fix them by the requirement that the deconvolved distribution vanishes on a tail set $T$, where the true ground-state weight is negligible. We obtain $T$ without the exact solution, from a WKB criterion,

$$T = \{j : e^{-2S_j} < 10^{-2}\}, \qquad S_j = \int_{y_{\min}}^{y_j} \sqrt{\frac{2\,[V(y) - E_{\text{ref}}]_+}{s_0^2}}\, dy, \tag{12}$$

where $S_j$ is the WKB action from the potential minimum $y_{\min}$ to the grid point $y_j$, $[x]_+ = \max(x, 0)$, and $E_{\text{ref}}$ is an energy estimate from the same run. We use the estimate after a uniform noise-floor subtraction. An overestimated $E_{\text{ref}}$ shrinks $T$, so the criterion is conservative. For all configurations of this work $T$ contains at least $n$ bins and its true weight is below $3.3 \times 10^{-4}$. We then solve

$$\min_{0 \leq \varepsilon_q \leq 0.2} \sum_{j \in T} \left[\left(\otimes_q B(\varepsilon_q)\right)^{-1} p^{\text{meas}}\right]_j^2 \tag{13}$$

where the sum runs over the tail set $T$ and the upper bound on $\varepsilon_q$ keeps the inverse well conditioned. Finally we project the deconvolved vector onto the simplex. As a simpler baseline we also report a uniform floor subtraction. It estimates the floor from the smallest bins and subtracts it.

The threshold $10^{-2}$ in Eq. (12) was fixed for all runs before the hardware data were analyzed. Both the deconvolution and the floor subtraction return a probability distribution, so that the bound $E[p] \geq E_0$ of Section 4.1 is preserved. Appendix D separates the contributions of the flip model and of the tail set. On synthetic data with known flip rates, the deconvolution recovers the flip rates of the two most significant qubits within 0.12 percentage points and reduces the energy error from 760–2,150 bp to 10–36 bp. Applied to the exact ground state, it changes the energy by less than 8 bp. On the hardware data, a plain truncation of the tail set $T$ followed by renormalization gives errors comparable to, or smaller than, those of the deconvolution. Hence most of the correction comes from the tail set, that is, from locating without the exact solution the bins where the true weight is negligible. The flip model does not consistently outperform a direct truncation of the tail set. The deconvolution uses the same tail set within a physical noise model. It returns flip rates that can be compared across runs and platforms, and it moves the removed weight back to the bins from which it leaked.

### 4.4 Zero-noise extrapolation

We amplify the two-qubit noise by CNOT folding, $\mathrm{CX} \rightarrow \mathrm{CX}\,(\mathrm{CX}\,\mathrm{CX})^{(\lambda-1)/2}$ with $\lambda = 1{,}3{,}5$ [12, 13]. Barriers between the folded gates prevent cancellation by the transpiler, and all folded circuits use the same physical qubits. The main result is the linear extrapolation through $\lambda = 1$ and 3. The quadratic extrapolation through $\lambda = 1$, 3 and 5 is reported as a check.

## 5. Hardware experiments

### 5.1 Setup

The experiments were performed on the IBM superconducting processor ibm_cleveland in two sessions on 17 October 2026 (KST), using the Qiskit Runtime Sampler (qiskit 2.5.2, qiskit-ibm-runtime 0.50.0). Gate twirling and XpXm dynamical decoupling were enabled. Measurement twirling was disabled to keep the readout calibration consistent. The ansatz parameters were trained classically (Section 3). The four-qubit circuits ran on the linear chain of physical qubits (24, 23, 22, 21) and the six-qubit circuit on (24, 23, 16, 3, 2, 1). After transpilation the circuits contained 9, 27 and 45 two-qubit gates for $L = 3$ at $\lambda = 1$, 3 and 5; 6, 18 and 30 for $L = 2$; and 30, 90 and 150 for the six-qubit circuit. The measured readout errors were 0.1–1.0% per qubit for the four-qubit chain and at most 1.3% for the six-qubit chain.

The first session had three stages. Stage A evaluated the four-qubit circuits for Vasicek ($L = 3$ and $L = 2$), CIR and Black–Karasinski with $10^5$ shots per circuit. Stage B repeated the four-qubit Vasicek circuit and evaluated the six-qubit Vasicek circuit with $10^6$ shots per circuit. Stage C trained the four-qubit Vasicek ansatz on the device (Section 5.4). Each readout calibration used $2 \times 10^4$ shots per circuit. The total QPU time was 69.8 min (A 6.2 min, B 27.6 min, C 35.9 min), which corresponds to about 274 μs per shot.

The second session started about 15 hours after the first. Stage D repeated the four-qubit circuits on the same physical qubits, together with two alternative encodings (Section 6). Stage E measured the same trained circuits with a standard Pauli-measurement estimator. Stage F repeated the binary circuits once more with automatic qubit selection. The second session used 34.0 min of QPU time (D 13.9 min, E 10.8 min, F 9.3 min). A third session on 25 October 2026 evaluated circuits with parameters calibrated to the U.S. Treasury curve (Section 5.6).

### 5.2 Trained-circuit evaluation

Table 2 lists the energy errors after each step of the analysis. Readout mitigation alone leaves errors of 566–1,139 bp at four qubits. Since the readout errors are small, the remaining error is due mainly to gate noise. The uniform floor subtraction combined with zero-noise extrapolation reduces the error to 125–330 bp. The bit-flip deconvolution combined with zero-noise extrapolation reduces it further to 33–115 bp, which is a relative error of 0.7–2.5%. At the same time the distribution fidelity increases from 0.962–0.980 to 0.987–0.996.

**Table 2.** Hardware results on ibm_cleveland, session 1. Energy errors $E - E_{\mathrm{grid}}$ are in basis points; fidelity is the Bhattacharyya overlap with $p_0$ at $\lambda = 1$. ZNE is linear through $\lambda = 1$ and 3.

| Configuration | Shots per circuit | Readout-mitigated | Floor subtracted | Floor + ZNE | Deconvolved | **Deconvolved + ZNE** | Fidelity (mitigated → deconvolved) |
|---|---|---|---|---|---|---|---|
| CIR, $n = 4, L = 3$ | $10^5$ | +686 | +273 | +125 | +56 | **+33** | 0.980 → 0.996 |

| Configuration | Shots per circuit | Readout-mitigated | Floor subtracted | Floor + ZNE | Deconvolved | **Deconvolved + ZNE** | Fidelity (mitigated → deconvolved) |
|---|---|---|---|---|---|---|---|
| Vasicek, $n = 4$, $L = 2$ | $10^5$ | +566 | +285 | +200 | +89 | **+43** | 0.977 → 0.993 |
| Vasicek, $n = 4$, $L = 3$ | $10^6$ | +788 | +328 | +214 | +133 | **+89** | 0.966 → 0.990 |
| Vasicek, $n = 4$, $L = 3$ | $10^5$ | +704 | +293 | +179 | +144 | **+101** | 0.967 → 0.989 |
| Black–Karasinski, $n = 4, L = 3$ | $10^5$ | +1,139 | +449 | +330 | +118 | **+115** | 0.962 → 0.987 |
| Vasicek, $n = 6$, $L = 6$ | $10^6$ | +1,388 | +1,080 | +403 | +971 | +518 | 0.962 → 0.987 |

Figure 3a shows the origin of the residual error for the Vasicek circuit with $10^6$ shots. The leaked probability is not uniform. It is concentrated on single-bit-flip neighbors of the peak states $j = 7$ (0111) and $j = 8$ (1000), in particular on $j = 15$ and $j = 0$, which are reached by a flip of the most significant qubit. The fitted flip rates are $\boldsymbol{\varepsilon} = (0.0165, 0.0125, 0, 0)$ from the most to the least significant qubit. The deconvolution removes the edge weight while it leaves the central bins unchanged. The fidelity extrapolated to $\lambda = 0$ is 0.982 before deconvolution. This indicates a noise component that does not scale with CNOT folding, which explains why zero-noise extrapolation alone cannot remove the bias.

Figure 3b shows the deconvolved energy versus $\lambda$. For Black–Karasinski the dependence on $\lambda$ is weak (118, 123 and 133 bp), whereas for the other circuits it is close to linear between $\lambda = 1$ and 3. For the six-qubit circuit the deconvolution reduces the error only from 1,080 to 971 bp at $\lambda = 1$. At $n = 6$ the kinetic coefficient $s_0^2/(2h^2)$ is 18 times larger than at $n = 4$. Hence local roughness of $\sqrt{p}$ between neighboring bins is strongly amplified, and this roughness is not of the bit-flip form. On the other hand, a plain truncation of the WKB tail set, which assumes no noise model, reduces the error to 216 bp at $\lambda = 1$ and to 77 bp after extrapolation (Appendix D). The true ground-state weight in this tail set is $5 \times 10^{-5}$.

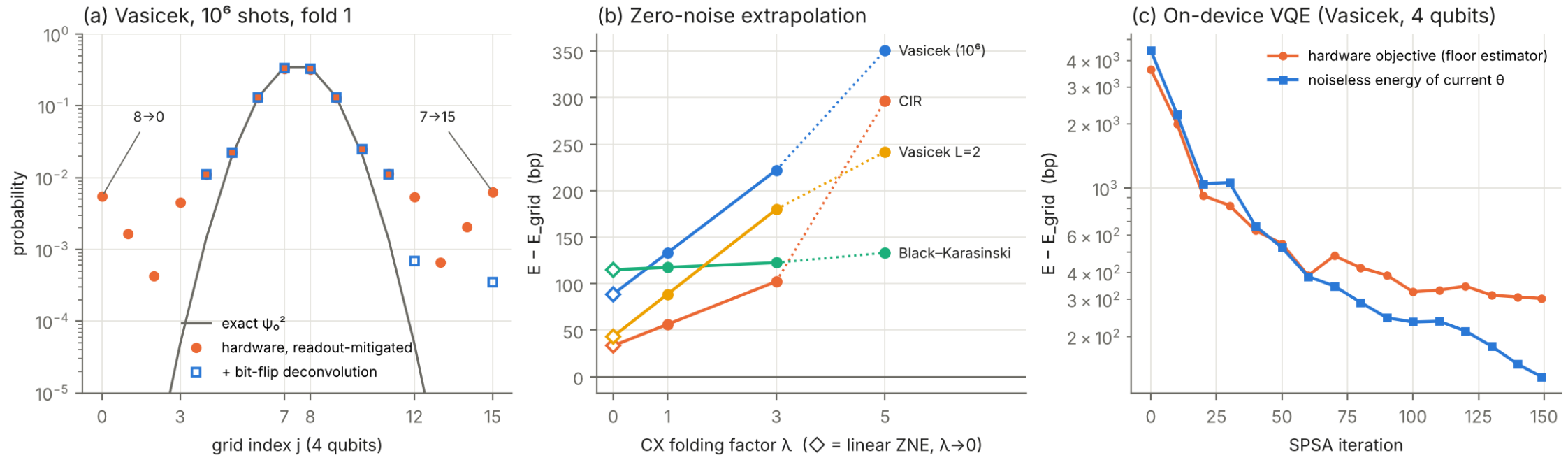

***Figure 3.*** *Hardware results on ibm_cleveland. (a) Vasicek, four qubits,* $10^6$ *shots,* $\lambda = 1$*: exact* $p_0$ *(line), readout-mitigated distribution (filled circles) and bit-flip-deconvolved distribution (open squares), on a logarithmic scale. Arrows mark the leakage* $7 \rightarrow 15$ *and* $8 \rightarrow 0$ *by a flip of the most significant qubit. (b) Deconvolved energy error versus the CNOT folding factor* $\lambda$ *for the four-qubit circuits; open diamonds are the linear extrapolations to* $\lambda = 0$*. (c) On-device SPSA training of the four-qubit Vasicek ansatz: hardware objective and noiseless energy of the current parameters.*

### 5.3 Reproducibility and comparison with Pauli-measurement VQE

Table 3 collects the deconvolved and extrapolated errors of the binary-encoded four-qubit circuits from all runs. Over ten runs in two sessions the errors range from 33 to 125 bp, with a mean of 90 bp. The Black–Karasinski result is the most stable (111–125 bp). The CIR result varies most (33–120 bp).

**Table 3.** Reproducibility of the deconvolved and extrapolated energy error (bp) for the binary-encoded four-qubit circuits. Shots per circuit are $10^5$ unless noted.

| Model | Session 1 | Session 2, same qubits | Session 2, automatic qubit selection | Range |
|---|---|---|---|---|
| Vasicek | +101; +89 ($10^6$ shots) | +96 | +44 | 44–101 |
| CIR | +33 | +69 | +120 | 33–120 |
| Black–Karasinski | +115 | +111 | +125 | 111–125 |

In stage E we measured the same trained circuits, on the same physical qubits and in the same session, with the Qiskit Runtime Estimator. The Hamiltonian was decomposed into Pauli strings and grouped into nine qubit-wise commuting groups for each model. Each group was measured with $2 \times 10^4$ shots, with TREX readout mitigation, and in a second run with TREX and zero-noise extrapolation (noise factors 1, 3 and 5, linear extrapolation). Table 4 compares the results.

**Table 4.** Pauli-measurement VQE versus the $\sqrt{p}$ estimator on the same circuits (session 2, ibm_cleveland). Energy errors in bp; ± values are the standard errors reported by the Estimator. The two pipelines used different shot budgets and mitigation protocols (see text), and the comparison refers to these conditions.

| Model | Pauli, TREX | Pauli, TREX + ZNE | $\sqrt{p}$, readout-mitigated | $\sqrt{p}$, deconvolved + ZNE |
|---|---|---|---|---|
| Vasicek | +1,290 ± 55 | +791 ± 164 | +759 | **+96** |
| CIR | +1,525 ± 158 | +1,129 ± 475 | +663 | **+69** |
| Black–Karasinski | +1,499 ± 75 | +959 ± 233 | +1,124 | **+111** |

The readout-mitigated $\sqrt{p}$ estimator alone is already closer to $E_0$ than the Pauli estimator with TREX. With deconvolution and extrapolation, the error of the $\sqrt{p}$ pipeline is about ten times smaller than that of the Pauli estimator with TREX and ZNE. The two pipelines do not use identical resources. The Estimator runs used $2 \times 10^4$ shots each with the built-in TREX and ZNE of Qiskit Runtime, whereas the $\sqrt{p}$ values use $10^5$ shots per folding factor and the analysis of Section 4. The statistical errors of the Estimator (55–475 bp) are much smaller than its bias, so the difference is not a sampling effect. The factor of ten therefore refers to these experimental conditions. It also costs less: one measurement setting per circuit instead of nine, and 1.5 min of QPU time per model including folding and readout calibration, against 3.6 min for the two Estimator runs. The advantage follows from Section 4.1. The Pauli estimator sums many terms with large and partly cancelling coefficients, whereas the $\sqrt{p}$ estimator uses one distribution, and its errors can be traced to leakage into identifiable bins.

### 5.4 On-device VQE

Following the hardware-efficient VQE of Kandala et al. [16], we trained the four-qubit Vasicek ansatz on the device in stage C, using SPSA [14] with the $\sqrt{p}$ estimator as the cost function. Specifically, the cost was the floor-subtracted $\sqrt{p}$ energy of the readout-mitigated distribution at $\lambda = 1$. Each SPSA iteration evaluated the two perturbed parameter sets in one job with $2 \times 10^4$ shots each. We used the standard gain sequences with exponents 0.602 and 0.101, a perturbation size of 0.15 rad and a first step calibrated to 0.15 rad. The readout was recalibrated every 50 iterations. The initial parameters were the classically trained values plus Gaussian perturbations with a standard deviation of 0.3 rad, which gives a noiseless energy error of 4,939 bp and a fidelity of 0.722.

After 150 iterations (172 jobs, 35.9 min of QPU time) the noiseless energy error of the trained parameters is 130 bp and the fidelity is 0.987 (Fig. 3c). The hardware objective decreases from 3,632 bp to 303 bp. The difference between the two curves is the noise bias of the hardware cost function. The bias does not prevent convergence, since it changes slowly with the parameters. The logical and compiled circuits and the parameters before and after training are given in Appendix F.

### 5.5 Cross-platform check on a trapped-ion processor

In order to test whether the results depend on the hardware platform, we evaluated the trained four-qubit Vasicek circuit on the IonQ trapped-ion processor Forte Enterprise (qpu.forte-enterprise-1) on 18 October 2026 (KST). The circuit was the same binary-encoded $L = 3$ circuit used on ibm_cleveland, with nine CNOT gates. It was compiled to the IonQ QIS gate set without optimization (qiskit-ionq 1.1.1) and submitted through the IonQ Cloud API with debiasing disabled, as on IBM. We did not use zero-noise extrapolation, since the IonQ compiler may cancel folded gate pairs. The comparison is therefore made at $\lambda = 1$. We performed two runs, with 500 and 5,000 shots for the circuit. Each run had its own readout calibration with $|0000\rangle$ and $|1111\rangle$ circuits of 500 shots each.

The analysis used the raw shot histograms returned by the IonQ API. For the calibration circuits, the probability vectors returned by the same API omitted the low-weight outcomes, and their sums were 0.990 and 0.968 in the two runs. The histograms contain all shots. A histogram gives the number of shots of each outcome, keyed by the integer value of the bit string with qubit 0 as the least significant bit. We converted it to the bit order of Section 2.5 before the analysis. The $|0000\rangle$ calibration showed no error in either run. For $|1111\rangle$ the per-qubit readout errors were 0.2–0.4% in the first run and 0.4–1.0% in the second run, comparable to 0.1–1.0% on ibm_cleveland. A $\chi^2$ test of homogeneity on the 16 outcome counts of the two runs (15 degrees of freedom) gives $\chi^2 = 14.1$ and $p = 0.52$. Hence the test detects no difference between the two runs; this does not establish that the underlying distributions are equal. The bit-flip rates fitted by the deconvolution in the second run are 2.2% and 1.4% for the two most significant qubits, against 1.7% and 1.3% on ibm_cleveland.

**Table 5.** Vasicek four-qubit circuit on IonQ Forte Enterprise and ibm_cleveland at $\lambda = 1$, after readout mitigation. Energy errors $E - E_{\text{grid}}$ in bp. For IonQ, the ± values are standard deviations over 400 bootstrap replicates, in which the circuit histogram and the $|1111\rangle$ calibration histogram are resampled multinomially.

| Platform, run, shots | $\sqrt{p}$ | Floor-subtracted | Deconvolved | Fidelity |
|---|---|---|---|---|
| IonQ Forte Enterprise, run 1, 500 | +935 ± 243 | +444 ± 184 | +246 ± 107 | 0.954 |
| IonQ Forte Enterprise, run 2, 5,000 | +1,021 ± 117 | +420 ± 95 | +254 ± 53 | 0.956 |
| ibm_cleveland, session 1, $10^5$ | +704 | +293 | +144 | 0.967 |
| ibm_cleveland, session 1, $10^6$ | +788 | +328 | +133 | 0.966 |

On both processors the analysis reduces the error in the same proportions. The floor subtraction removes about 60% of the readout-mitigated error on both platforms. The deconvolution leaves 25% of it on IonQ and 17–20% on ibm_cleveland. The readout-mitigated IonQ error of run 2 is larger by about 250 bp, which is about two standard deviations. The deconvolved errors differ by about 115 bp, also about two standard deviations. Hence the bias structure of Section 5.2, which is leakage by effective flips of the high-order qubits, appears on both platforms with similar flip rates. We do not attribute the residual difference to the hardware. The IonQ runs used 500 and 5,000 shots with 500-shot readout calibrations, whereas the IBM runs used $10^5$–$10^6$ shots with $2 \times 10^4$-shot calibrations. A difference of two standard deviations at these sample sizes is not sufficient to rank the two processors.

### 5.6 Treasury-calibrated circuits

In order to connect the hardware pipeline with market data, we repeated the evaluation of Section 5.2 with parameters calibrated to the U.S. Treasury curve of 16 October 2026 (Section 2.4 and Table E1). The mean-reversion speed was bounded by $\kappa \geq 0.1$. The volatility of the Vasicek model was fixed at the realized volatility of the 1-year yield over the preceding year, 53 bp per year, and the CIR and Black–Karasinski volatilities were chosen to give the same absolute volatility at the mean short rate of that year (3.86%). The calibrated parameters are $\kappa = 0.153$, $\theta = 5.97\%$ and $\sigma = 0.53\%$ for Vasicek; $\kappa = 0.149$, $\theta = 6.01\%$ and $\sigma = 0.0271$ for CIR; and $\kappa = 0.159$, $\bar{r} = 5.85\%$ and $\sigma = 0.138$ for Black–Karasinski. All three models fit the 1–30 year yields with an error of 6 bp and give a continuum long-term yield of 5.90–5.91%, against a 30-year zero yield of 5.46%. The four-qubit circuits ($L = 3$, nine CNOT gates) were trained classically and reproduce the grid ground states with a fidelity of 1.0000. They were evaluated in a third session on 25 October 2026 (KST), on the same physical qubits (24, 23, 22, 21) and with the same settings as stage A (qiskit 2.6.1, qiskit-ibm-runtime 0.51.0). The three models used 4.7 min of QPU time.

Table 6 lists the results. As in the rest of Section 5, the errors refer to the four-qubit grid ground-state energies, not to market yields or to the continuum long-term yields. The analysis reduces the error from 86–657 bp after readout mitigation to 23–67 bp after deconvolution and extrapolation. These values lie at the lower end of the range of Table 3 (33–125 bp). Part of the improvement is due to the parameters. The calibrated volatilities are smaller than those of Section 3, and the energy shift caused by a 1% flip of the most significant qubit is reduced from 358–557 bp to 110–297 bp. The CIR circuit is the least sensitive (110 bp), and its readout-mitigated error is accordingly small (86 bp). For Vasicek the fitted flip rates of the two most significant qubits are 2.1% and 3.2%, compared with 1.7% and 1.3% in session 1. The bias therefore has the same structure as in Section 5.2.

**Table 6.** Treasury-calibrated four-qubit circuits on ibm_cleveland (session 3, 25 October 2026, $10^5$ shots per circuit). $E_{\text{grid}}$ and $E_\infty$ are the grid and continuum ground-state energies. Energy errors $E - E_{\text{grid}}$ in bp; fidelity at $\lambda = 1$. ZNE is linear through $\lambda = 1$ and 3.

| Model | $E_{\text{grid}}$ (%) | $E_\infty$ (%) | Readout-mitigated | Floor + ZNE | Deconvolved | **Deconvolved + ZNE** | Fidelity (mitigated → deconvolved) |
|---|---|---|---|---|---|---|---|
| Vasicek | 5.692 | 5.911 | +657 | +45 | +134 | **+67** | 0.942 → 0.987 |
| CIR | 5.801 | 5.911 | +86 | 0 | +46 | **+23** | 0.989 → 0.993 |
| Black–Karasinski | 5.550 | 5.898 | +448 | +152 | +67 | **+63** | 0.960 → 0.986 |

The hardware estimates of the long-term yield are 6.37%, 6.03% and 6.18% for the three models. The grid error at $n = 4$ is $-11$ to $-35$ bp and has the opposite sign to the hardware bias. Hence the hardware estimates differ from the continuum values by only 12–45 bp. We do not rely on this partial cancellation, since the two errors have different origins. On the other hand, the hardware error is of the same order as the spread of $E_0$ among the three models on the dates of Table E1 and as the calibration uncertainty discussed in Appendix E. The complete chain from the market curve to a hardware estimate of the long-term yield thus works with the present four-qubit circuits.

## 6. Discussion

The long-term yield is a natural target for ground-state algorithms for three reasons. First, it is exactly a ground-state energy, not a property of a time-evolved state. Second, the Feynman–Kac Hamiltonian is stoquastic, so that a single computational-basis measurement setting suffices. Third, the estimator $E[p]$ is a variational upper bound for any measured distribution. Readout mitigation with simplex projection and the bit-flip deconvolution both return distributions, and therefore they preserve this bound. Only the zero-noise extrapolation can cross below $E_0$.

The hardware data show that the structure of the noise matters more than its magnitude. For the binary-encoded circuits, the leaked weight is described by effective bit flips of the high-order qubits. The fitted flip rates reproduce the measured bias: 727, 613 and 1,090 bp predicted against 759, 663 and 1,124 bp measured for Vasicek, CIR and Black–Karasinski in session 2.

We tested in stage D whether this structure can be removed at the level of the encoding. In a Gray encoding, a flip of the most significant qubit reflects the grid about its center and maps the peak onto itself. We also optimized the grid-to-bitstring permutation to minimize the first-order bias $\sum_q \varepsilon\, S_q$ for a uniform flip rate $\varepsilon = 1\%$, where $S_q$ is the energy sensitivity of qubit $q$ derived from the exact identity

$$E[p] - E_0 = \langle \delta\psi \,|\, H - E_0 \,|\, \delta\psi \rangle, \qquad \delta\psi = \sqrt{p} - \sqrt{p_0}. \tag{14}$$

Here $\delta\psi$ is the deviation of the measured amplitude from the exact ground-state amplitude. Since $H - E_0$ is positive semidefinite, Eq. (14) also shows directly that $E[p] \geq E_0$. The sensitivity $S_q$ follows by expanding $\delta\psi$ to first order in the flip rate of qubit $q$. Both encodings keep $H$ stoquastic, and the three-layer ansatz with nine CNOT gates still prepares their ground states exactly. On hardware, the Gray encoding improved the floor-subtracted and extrapolated errors for Vasicek (+100 versus +195 bp) and CIR (−30 versus +130 bp), but not for Black–Karasinski (+244 versus +183 bp). The optimized encoding was worse than the binary one for all three models. The total leaked weight did not decrease, and the flip model fitted to one encoding did not predict the bias of the others. We conclude that the effective flip structure belongs to the trained circuit rather than to the device alone. A noise-aware encoding therefore requires a circuit-level noise model, which we leave for future work. A higher-order or spectral kinetic operator would reduce the number of qubits for a given discretization error, but it introduces positive off-diagonal elements and breaks the stoquastic structure.

The spectrum of $H$ carries more information than $E_0$. The gap $E_1 - E_0$ sets the rate at which the forward curve approaches $y_\infty$, and the low-lying states give the spectral expansion of the term structure [6]. These states can be obtained with excited-state extensions of the VQE. The same Hamiltonian also governs long-dated path-dependent quantities. For a functional $\int_0^\tau v\,(y_t)\,dt$ with $v \geq 0$, such as the occupation time of a range of rates, the Laplace-transformed price leads to the operator $H + s\,v$ [19, 21]. This operator is stoquastic for $s \geq 0$, and therefore the estimator of Section 4 applies without change (Appendix B). In contrast, the direct partial differential equation for the arithmetic Asian option has a time-dependent and non-Hermitian generator. Its solution is then not a projection onto a ground state [3, 26].

For one-factor models a classical eigensolver on a fine grid is inexpensive, and we use it as the reference. Calibrated to the U.S. Treasury zero curve from 1985 to 2026, the ground-state energy $E_0$ follows the 30-year yield with an in-sample correlation of 0.98 and lies above it by about 50 bp (Section 2.4). The motivation for the quantum formulation is the multi-factor case. A $d$-factor model requires $d \cdot n$ qubits, while a classical grid requires $2^{dn}$ points. In $d$ dimensions the diffusion term is $\frac{1}{2} g^{\mu\nu}\,\partial_\mu\,\partial_\nu$ with the metric $g^{\mu\nu} = \sum_a \sigma_a^\mu \sigma_a^\nu$ [21]. A global Lamperti transform to a constant diffusion requires, in particular, that this metric be flat [22]. Otherwise the kinetic operator becomes a Laplace–Beltrami operator with curvature corrections. Market data also show fat tails and jumps that a one-factor diffusion does not capture [25]. A jump component adds a non-local term to the generator. Since jump rates are non-negative, the corresponding off-diagonal elements of $H$ are non-positive, and $H$ remains stoquastic. It is Hermitian after the similarity transform when the jump kernel satisfies detailed balance. In that case the Hamiltonian is no longer tridiagonal, but the estimator of Section 4 still applies with the additional off-diagonal terms. Non-affine multi-factor models, such as shadow-rate models with a zero lower bound, have no closed form. When the drift is reversible, their generators are symmetrizable in the same way as in Section 2, and the Hamiltonian remains stoquastic. In that regime the ground state is concentrated on a small part of the grid. Sample-based quantum diagonalization [15], which diagonalizes $H$ in the sampled support, then becomes a natural complement to the estimator of Section 4.

## 7. Limitations

1. All models in this work are one-factor models, for which classical solvers are inexpensive. The hardware results demonstrate the method; they do not claim a computational advantage.
2. The hardware errors are measured with respect to the ground state of the same grid. At four qubits the discretization error with respect to the continuum (−22.5 to −64.9 bp) is comparable to the hardware error. A quantitative continuum result requires more qubits or a higher-order discretization.
3. At six qubits the bit-flip deconvolution gives 518 bp after extrapolation, since the leaked weight is not of the bit-flip form. A plain truncation of the tail set gives 77 bp after zero-noise extrapolation (216 bp at $\lambda = 1$), but it rests on no noise model, and its accuracy at larger $n$ remains to be established.
4. The tail set relies on the design condition that the grid is wide enough for the ground state to vanish near its edges. The WKB criterion verifies this condition from the potential, but it is approximate. The corrected energy depends on the threshold of Eq. (12); thresholds of $10^{-3}$ and $10^{-1}$ give larger errors than $10^{-2}$ (Appendix D).
5. The bit-flip model is phenomenological. It captures the dominant leakage of the binary-encoded circuits, but it is specific to the trained circuit and does not transfer to other encodings (Section 6). It does not describe coherent or correlated errors. The linear zero-noise extrapolation assumes that the bias is linear in the folding factor between $\lambda = 1$ and 3.
6. The superconducting data come from one processor in three sessions, including the Treasury-calibrated evaluation. The trapped-ion check covers one model at $\lambda = 1$ without zero-noise extrapolation, with readout calibrations of 500 shots.
7. The extension to multi-factor models requires a symmetrizable, that is reversible, drift. This condition must be checked for each calibrated model.
8. The Treasury calibration is in sample. The calibrated long-term yield depends on the one-factor model, on the bound on $\kappa$ and on the extrapolation beyond 30 years, and the Black–Karasinski fit is unreliable on some near-zero-rate dates (Appendix E).

## 8. Conclusion

We have shown that the long-term yield of a short-rate model is the ground-state energy of a stoquastic Schrödinger operator obtained from the Feynman–Kac equation. This makes the long-term yield a natural target for the VQE. In noiseless simulation, a six-qubit real-amplitude ansatz reproduces the grid ground state within 0.06 bp for the Vasicek model, and the same formulation reproduces the closed-form CIR yield and gives the Black–Karasinski yield, for which no closed form exists. Since the Hamiltonian is stoquastic, the energy can be estimated from computational-basis samples alone. On ibm_cleveland we identified bit flips of high-order qubits as the dominant source of bias and removed most of it by a tail-constrained bit-flip deconvolution. The four-qubit hardware estimates agree with the grid ground-state energies within 33–125 bp for all three models in ten runs over two sessions. On a trapped-ion processor the same pipeline reduces the error of the Vasicek circuit from 1,021 to 254 bp at $\lambda = 1$, in the same proportions as on ibm_cleveland, which shows that the bias structure is not specific to one platform. On the same circuits and under the conditions of Section 5.3, its error is about ten times smaller than that of a Pauli-measurement VQE with readout mitigation and zero-noise extrapolation, at a lower measurement cost. At six qubits a truncation of the WKB tail set gives 77 bp after zero-noise extrapolation. An on-device VQE converges from 4,939 bp to 130 bp. Calibrated to the U.S. Treasury curve over 41 years, the long-term yield $E_0$ follows the 30-year market yield closely. With the parameters calibrated to the curve of October 2026, the hardware estimates agree with the grid values within 23–67 bp, so that the complete chain from the market curve to a hardware estimate of the long-term yield has been demonstrated. The extension to multi-factor and non-affine models, together with sample-based quantum diagonalization, is the subject of future work.

## Acknowledgments

This work was supported by Korea National Research Foundation (NRF) grant No. NRF-2026-25612745, ICT R&D program of IITP-2023-2021-0-01810, RS-2023-00225385, AFOSR grant FA2386-22-1-4052. The work was partly supported by the Cleveland Clinic Quantum Start-up Catalyzer Program and Amazon Web Services. The author acknowledges the use of IBM Quantum services for this work through the Cleveland Clinic's Quantum Start-up Catalyzer Program. The views expressed are those of the author, and do not reflect the official policy or position of IBM or the IBM Quantum team. This work was also supported by the National Quantum Laboratory at Maryland (QLab). The author acknowledges QLab for providing quantum computing resources on IonQ devices for this work. The views expressed are those of the author, and do not reflect the official policy or position of IonQ. D. Ahn is a co-founder of Singularity Quantum Inc. and holds an equity interest in the company.

## Data availability

Raw IBM counts of all three sessions, raw IonQ shot histograms, job identifiers, trained parameters and the analysis code will be made publicly available in Figshare with a DOI upon publication. The repository will also contain the compiled circuit of every IBM job of sessions 1 and 2 and the submitted circuit of every IonQ job, together with scripts that reproduce the hardware tables.

## Declaration of generative AI and AI-assisted technologies in the manuscript preparation process

During the preparation of this work, the author used Claude (Anthropic) and ChatGPT (OpenAI) to assist in writing the analysis and job-submission code, in the data analysis, and in drafting and editing the text. After using these tools, the author reviewed and edited the content as needed and takes full responsibility for the content of the published article.

## Appendix A. Path-integral derivation of the Feynman–Kac Hamiltonian

In this appendix we derive Eq. (3) from the discretized path integral, following the constructions of Baaquie [19] and Linetsky [21]. In the Lamperti coordinate the short rate follows $dy_t = \tilde{b}(y_t)\, dt + s_0\, dW_t$. We divide the interval $[0,\tau]$ into $N$ steps of length $\Delta t = \tau/N$ and write $y_i = y(i\Delta t)$ with $y_0 = y$. In the pre-point (Itô) discretization the short-time transition density is Gaussian. The bond price is then the limit of a finite-dimensional integral,

$$P(\tau, y) = \lim_{N\to\infty} \int \prod_{i=1}^{N} \frac{dy_i}{\sqrt{2\pi s_0^2 \Delta t}}\; e^{-S_N}, \qquad S_N = \sum_{i=0}^{N-1} \left[ \frac{\left(\Delta y_i - \tilde{b}(y_i)\,\Delta t\right)^2}{2 s_0^2 \Delta t} + r(y_i)\,\Delta t \right], \tag{A1}$$

where $\Delta y_i = y_{i+1} - y_i$. In the limit $N \to \infty$ the sum $S_N$ becomes the Euclidean action of Eq. (5). By expanding the square, we obtain

$$S_N = \sum_{i=0}^{N-1} \left[ \frac{(\Delta y_i)^2}{2 s_0^2 \Delta t} + \left( r(y_i) + \frac{\tilde{b}(y_i)^2}{2 s_0^2} \right) \Delta t \right] - \sum_{i=0}^{N-1} \frac{\tilde{b}(y_i)}{s_0^2}\, \Delta y_i. \tag{A2}$$

The last sum is a stochastic integral in the pre-point form. In order to evaluate it, we expand $W(y) = \int \tilde{b}/s_0^2\, dy$ to second order and use the Itô rule $(\Delta y_i)^2 \to s_0^2 \Delta t$. Then

$$W(y_{i+1}) - W(y_i) = \frac{\tilde{b}(y_i)}{s_0^2}\, \Delta y_i + \frac{\tilde{b}'(y_i)}{2}\, \Delta t, \tag{A3}$$

and the last sum equals $W(y_N) - W(y_0) - \sum_i \tilde{b}\,'(y_i)\, \Delta t/2$. As a result,

$$S_N = \sum_{i=0}^{N-1} \left[ \frac{(\Delta y_i)^2}{2 s_0^2 \Delta t} + V(y_i)\, \Delta t \right] - W(y_N) + W(y_0), \tag{A4}$$

where $V$ is the potential of Eq. (3). Here the term $\tilde{b}'/2$ appears as the Itô correction of the pre-point rule. A mid-point (Stratonovich) discretization gives the same Hamiltonian. In that case the stochastic integral has no correction term, and the same term arises from the normalization of the short-time propagator [21]. The remaining path integral is the Euclidean propagator $K(y', \tau \,|\, y) = \langle y' | e^{-H\tau} | y \rangle$ of the operator $H$ of Eq. (3). Therefore

$$P(\tau, y) = e^{-W(y)} \int dy'\, e^{W(y')}\, K(y', \tau \,|\, y), \tag{A5}$$

which is Eq. (4) after the expansion of $K$ in the eigenstates of $H$.

Two remarks connect this result with familiar cases. First, the kinetic term $(\Delta y_i)^2/(2 s_0^2 \Delta t)$ is the action $\frac{1}{2} m \dot{y}^2 \Delta t$ of a free particle with mass $m = 1/s_0^2$ (with $\hbar = 1$). Hence $H$ is the Schrödinger operator of a particle of mass $1/s_0^2$ in the potential $V$. Baaquie made this identification for the Black–Scholes equation, with mass $1/\sigma^2$ [23]. Second, for a constant drift $\tilde{b} = \mu$ and a constant rate $r$, we have $W = \mu y/s_0^2$ and $V = r + \mu^2/(2 s_0^2)$. The action then separates into the free action, the boundary term $-\mu(y_N - y_0)/s_0^2$ and the constant $\mu^2 \tau/(2 s_0^2)$. This is the decomposition of the Black–Scholes action given in Ref. [21]. The similarity transform of Section 2.2 is the generalization of this decomposition to a state-dependent drift.

## Appendix B. Long-dated path-dependent functionals

We consider a functional $I_\tau = \int_0^\tau v\,(y_t)\,dt$ with $v \geq 0$. Examples are the occupation time of a range of rates, $v = \mathbf{1}_{[a,b]}$, and the time-averaged rate. For $s \geq 0$ the Laplace transform

$$L(\tau, y; s) = \mathbb{E}\big[\exp\!\big(-\int_0^\tau r_t\;dt - s\,I_\tau\big)\,\big|\,y_0 = y\big] \tag{B1}$$

is a bond price in which the short rate is replaced by $r + s\,v$ [19, 21]. Prices of claims that depend on $I_\tau$ are obtained from $L$ by an inverse Laplace transform. By the derivation of Section 2, $L$ is governed by the operator $H(s) = H + s\,v$, and

$$\lim_{\tau\to\infty}\left[-\frac{1}{\tau}\ln L(\tau, y; s)\right] = E_0(s), \tag{B2}$$

where $E_0(s)$ is the ground-state energy of $H(s)$. The function $E_0(s)$ determines the long-maturity behavior of the functional. In particular, $E_0(0) = y_\infty$, and by the Hellmann–Feynman theorem $E_0{}'(0) = \int v\,\phi_0^2\,dy$ is the long-run mean of $v$ under the measure defined by the ground state.

The perturbation $s\,v$ is diagonal and real. Hence $H(s)$ has the same off-diagonal elements as $H$, and it remains stoquastic for every $s \geq 0$. The ground state of $H(s)$ is positive, and $E_0(s)$ can be estimated from computational-basis samples by Eq. (10) with $d_j$ replaced by $d_j + s\,v(y_j)$. On the other hand, the Fourier representation used for functionals that are not positive leads to an imaginary potential. The resulting operator is not Hermitian, and it lies outside the scope of the present method.

## Appendix C. Proof of Theorem 1

Let $\psi$ be a normalized eigenvector of $H$ with the eigenvalue $E_0$, and let $|\psi|$ denote the vector with the components $|\psi_j|$. Since $H_{jk} \leq 0$ for $j \neq k$, we have $H_{jk}\psi_j\psi_k \geq H_{jk}|\psi_j||\psi_k|$. Therefore

$$\begin{aligned}\langle|\psi|\,|\,H\,|\,|\psi|\rangle = \sum_j H_{jj}\;|\psi_j|^2 + \sum_{j\neq k} H_{jk}\;|\psi_j|\,|\psi_k| \;\leq\; \sum_j H_{jj}\;\psi_j^2 + \sum_{j\neq k} H_{jk}\;\psi_j\psi_k \\ = E_0.\end{aligned} \tag{C1}$$

Since $|\psi|$ is normalized, the variational principle gives $\langle|\psi|\,|\,H\,|\,|\psi|\rangle \geq E_0$. Hence the equality holds, and $|\psi|$ is also an eigenvector with the eigenvalue $E_0$.

Next we show that $|\psi|$ has no zero component. Suppose that $|\psi_k| = 0$. The $k$-th component of $(H - E_0)|\psi| = 0$ gives

$$\sum_{j\neq k} H_{kj}\;|\psi_j| = 0. \tag{C2}$$

Each term is non-positive. Therefore $|\psi_j| = 0$ for every $j$ with $H_{kj} \neq 0$. By repeating this argument and using the irreducibility of $H$, we reach every grid point, and we obtain $|\psi| = 0$. This contradicts the normalization. Hence $|\psi_j| > 0$ for all $j$. The same argument applies to any eigenvector with the eigenvalue $E_0$ that has non-negative components.

Finally, we consider $\varphi = |\psi| - \psi$. It is a linear combination of eigenvectors with the eigenvalue $E_0$, and $\varphi_j \geq 0$ for all $j$. If $\varphi = 0$, then $\psi = |\psi|$ is strictly positive. If $\varphi \neq 0$, then $\varphi$ is an eigenvector with non-negative components, and therefore $\varphi_j > 0$ for all $j$. In this case $\psi_j < |\psi_j|$, that is, $\psi_j < 0$ for all $j$. As a result, every eigenvector with the eigenvalue $E_0$ has the same sign in all its components. Two such eigenvectors cannot be orthogonal, since their inner product is a sum of terms of one sign. Therefore $E_0$ is nondegenerate, and its eigenvector can be chosen with strictly positive components. This completes the proof.

For the tridiagonal Hamiltonian of Eq. (9), $H_{j,j+1} = e < 0$ for all $j$, and therefore $H$ is irreducible. The Gray and optimized encodings of Section 6 only permute the basis states. They preserve both the sign structure and the irreducibility, so that Theorem 1 applies to them as well.

## Appendix D. Validation and ablation of the tail-constrained deconvolution

In this appendix we separate the contributions of the bit-flip model and of the tail set. All energies are errors $E - E_{\text{grid}}$ in bp, and all tail sets use the threshold $10^{-2}$ of Eq. (12) unless stated otherwise.

**Synthetic data.** We apply independent bit flips with known rates $\boldsymbol{\varepsilon}$ to the exact grid ground state, add a uniform floor of 0.5%, and analyze the result as in Section 4. Table D1 lists the results for two sets of rates. The deconvolution recovers the flip rates of the two most significant qubits within 0.12 percentage points, and it reduces the energy error to 10–36 bp. For CIR it also assigns spurious rates of 0.6–3% to the low-order qubits, but these have little effect on the energy. Applied to the exact ground state without noise, the deconvolution changes the energy by 1.7, 2.9 and 7.8 bp for Vasicek, CIR and Black–Karasinski.

**Table D1.** Deconvolution of synthetic data. Rates in %, from the most to the least significant qubit.

| Model | $\boldsymbol{\varepsilon}$ (true) | Noisy | Floor-subtracted | Deconvolved | $\boldsymbol{\varepsilon}$ (fitted) |
|---|---|---:|---:|---:|---|
| Vasicek | (1.65, 1.25, 0, 0) | +811 | +381 | +15 | (1.75, 1.37, 0, 0) |
| Vasicek | (3, 2, 1, 0) | +1,377 | +666 | +24 | (3.10, 2.12, 0, 0) |
| CIR | (1.65, 1.25, 0, 0) | +760 | +107 | +10 | (1.76, 1.42, 2.07, 0.61) |
| CIR | (3, 2, 1, 0) | +1,318 | +245 | +13 | (3.10, 2.16, 3.16, 1.64) |
| Black–Karasinski | (1.65, 1.25, 0, 0) | +1,264 | +726 | +36 | (1.74, 1.34, 0.53, 0) |
| Black–Karasinski | (3, 2, 1, 0) | +2,154 | +1,258 | +36 | (3.09, 2.09, 1.52, 0) |

**Hardware data.** Table D2 compares, for the readout-mitigated distributions at $\lambda = 1$, the floor subtraction, a plain truncation of the tail set $T$ followed by renormalization, and the deconvolution with three thresholds. For the four-qubit circuits the truncation gives errors of 33–149 bp, comparable to or smaller than the deconvolution with the threshold $10^{-2}$ (56–144 bp). For the six-qubit circuit the difference is larger: the truncation gives 216 bp at $\lambda = 1$ and 77 bp after extrapolation, against 971 and 518 bp for the deconvolution. Hence the tail set carries most of the correction. The deconvolution is sensitive to the threshold. With $10^{-3}$ the tail set is too small, and for CIR it is empty, so that no correction is applied. With $10^{-1}$ the tail set contains bins with true weight, and the error rises to 87–381 bp at four qubits and 1,137 bp at six qubits. Combined with zero-noise extrapolation, the truncation gives −42 to +114 bp; for Black–Karasinski the extrapolated value falls below $E_0$, which is possible only through the extrapolation. We report the deconvolution as the main result because it rests on a noise model that is validated on synthetic data, it gives flip rates that can be compared across runs and platforms, and its extrapolated values remain above $E_0$ in all runs.

**Table D2.** Ablation on the hardware data (ibm_cleveland, binary encoding, $\lambda = 1$, readout-mitigated). Energy errors in bp.

| Session, configuration | Readout-mitigated | Floor-subtracted | Tail truncation | Deconvolved ($10^{-3}$) | Deconvolved ($10^{-2}$) | Deconvolved ($10^{-1}$) |
|---|---|---|---|---|---|---|
| 1, Vasicek | +704 | +293 | +149 | +200 | +144 | +375 |
| 1, Vasicek ($L = 2$) | +566 | +285 | +91 | +142 | +89 | +381 |
| 1, CIR | +686 | +273 | +61 | +686 | +56 | +87 |
| 1, Black–Karasinski | +1,139 | +449 | +39 | +318 | +117 | +302 |
| 2, Vasicek | +759 | +309 | +125 | +196 | +138 | +373 |
| 2, CIR | +663 | +282 | +63 | +663 | +97 | +115 |
| 2, Black–Karasinski | +1,124 | +358 | +33 | +367 | +117 | +371 |
| 1, Vasicek ($n = 6$, $10^6$ shots) | +1,388 | +1,080 | +216 | +738 | +971 | +1,137 |

## Appendix E. Details of the Treasury calibration

This appendix gives the details of the calibration of Section 2.4. We used the zero yields of the Gürkaynak–Sack–Wright curve [28], downloaded from the Federal Reserve Board on 24 October 2026. Days without a complete set of 1–30 year yields were removed, and for each month we used the last remaining business day, starting in November 1985, when the 30-year maturity becomes available. The last month is represented by 16 October 2026, the latest date available at download. The Vasicek and CIR parameters were fitted by least squares with the closed-form bond prices. The Black–Karasinski model was calibrated on eight selected dates, with the bond prices computed from the spectral expansion of Eq. (4).

A 30-year curve does not determine the mean-reversion speed well. Without a constraint, the fit drives $\kappa$ toward zero in about 16% of the months, and the Vasicek value $E_0 = \theta - \sigma^2/(2\kappa^2)$ then loses its meaning. We therefore bounded the mean-reversion speed by $\kappa \geq 0.1$, that is, a half-life of at most 7 years.

With this bound, $E_0$ follows the 30-year yield with a correlation of 0.98 for both models (Fig. 1). It lies above the 30-year yield by $46 \pm 38$ bp for Vasicek and $49 \pm 39$ bp for CIR (mean and standard deviation). Lower bounds $\kappa \geq 0.05$ and $\kappa \geq 0.2$ give mean offsets of 38–50 bp and correlations of 0.94–0.99. An offset of this size is expected, since the 30-year yield has not converged to $y_\infty$ and still contains the $1/\tau$ term of Eq. (4). The 30-year forward rate fluctuates more strongly, and $E_0$ exceeds it by $70 \pm 92$ bp (Vasicek). The gap $E_1 - E_0$ has a median of 0.23 per year for Vasicek and 0.25 per year for CIR. This corresponds to a time scale of about 4 years for the flattening of the forward curve.

Table E1 lists eight dates with different rate levels. The three models give values of $E_0$ within 50 bp of each other at most dates. At the near-zero short rates of 2012 and 2020, the Black–Karasinski model fits the curve poorly, with errors of 108 and 45 bp. In that case its value of $E_0$ is not reliable.

These values are the risk-neutral long-term yields implied by the curve and a one-factor model. They are not forecasts, and they depend on the extrapolation of the model beyond 30 years. On the other hand, the discount curves used in insurance regulation are extrapolated beyond the last liquid maturity toward a fixed ultimate forward rate. The ground-state energy $E_0$ of a calibrated model provides a model-consistent value for this quantity. The accuracy of the hardware estimates in Section 5 (33–125 bp, and 23–67 bp for the parameters calibrated to the curve of 16 October 2026 in Section 5.6) is of the same order as the model and calibration uncertainties found here.

**Table E1.** Calibration to the U.S. Treasury zero curve on selected dates ($\kappa \geq 0.1$). Rates in %; RMSE is the root-mean-square fitting error over the 1–30 year yields, in bp, for Vasicek / CIR / Black–Karasinski.

| Date | 1-y yield | 30-y yield | 30-y forward | $E_0$ Vasicek | $E_0$ CIR | $E_0$ Black–Karasinski | RMSE |
|---|---|---|---|---|---|---|---|
| 1985-11-29 | 7.79 | 10.65 | 11.36 | 11.01 | 10.98 | 10.97 | 5 / 5 / 6 |
| 1994-12-30 | 7.25 | 7.88 | 7.92 | 7.90 | 7.90 | 7.86 | 1 / 1 / 4 |
| 2000-01-31 | 6.29 | 6.26 | 5.41 | 5.51 | 6.21 | 5.82 | 3 / 7 / 4 |
| 2007-06-29 | 4.92 | 4.81 | 3.28 | 4.63 | 5.02 | 5.10 | 10 / 11 / 12 |
| 2012-07-31 | 0.18 | 2.67 | 2.47 | 3.83 | 4.10 | 2.79 | 12 / 16 / 108 |
| 2020-08-31 | 0.15 | 1.67 | 2.44 | 2.49 | 2.30 | 1.76 | 6 / 10 / 45 |
| 2023-10-31 | 5.32 | 5.06 | 4.40 | 5.29 | 5.29 | 5.18 | 12 / 12 / 13 |
| 2026-10-16 | 4.49 | 5.46 | 4.66 | 5.91 | 5.91 | 5.91 | 6 / 6 / 6 |

## Appendix F. Quantum circuit of the on-device VQE

This appendix gives the circuit used in the on-device VQE of Section 5.4. The same circuit, with the classically trained parameters, was used for the four-qubit Vasicek evaluations of Sections 5.2, 5.3 and 5.5.

Figure F1 shows the logical circuit. It is the real-amplitude ansatz of Section 3.1 with $n = 4$ and $L = 3$. Each layer applies $R_y(\theta_k) = \exp(-i\theta_k Y/2)$ to every qubit and then the CNOT ladder on the pairs $(q_0, q_1)$, $(q_1, q_2)$ and $(q_2, q_3)$. A final $R_y$ layer follows. The circuit has 16 parameters and 9 CNOT gates. Qubit $q_0$ is the most significant bit of the grid index $j$. For zero-noise extrapolation each CNOT is replaced by $\lambda$ CNOT gates in sequence, with $\lambda = 1$, 3 and 5. Since $\mathrm{CNOT}^2 = I$, the folded circuit implements the same unitary.

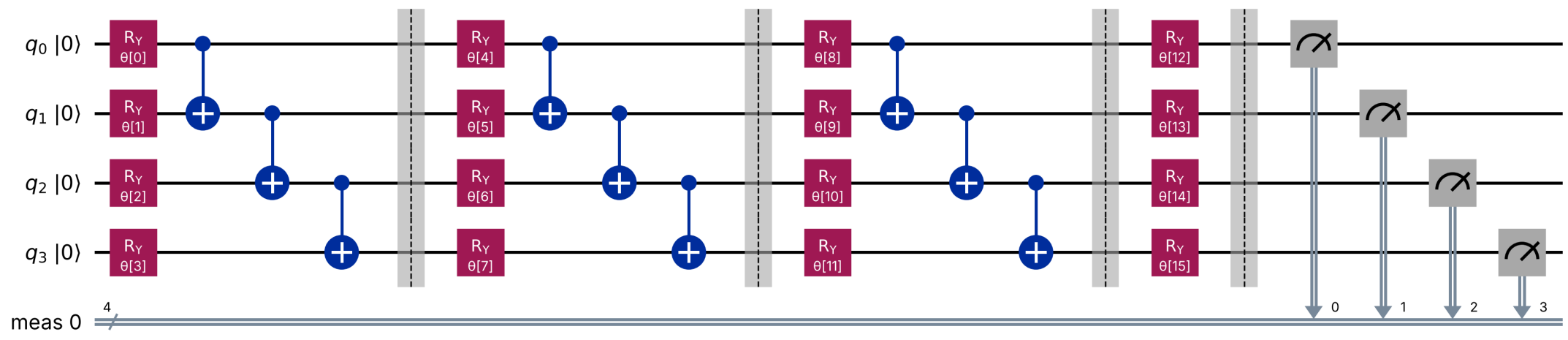

***Figure F1.*** *Logical circuit of the four-qubit ansatz ($n = 4$, $L = 3$) used in the on-device VQE. The barriers separate the layers. Qubit $q_0$ is the most significant bit of the grid index.*

Figure F2 shows the circuit after transpilation for ibm_cleveland with optimization level 3 and the fixed layout $(q_0, q_1, q_2, q_3) \rightarrow (24,23,22,21)$. The native gates are $R_z$, $\sqrt{X}$ and CZ. Each $R_y(\theta_k)$ becomes $\sqrt{X}\, R_z(\theta_k + \pi)\, \sqrt{X}$ up to fixed $R_z$ rotations, and each CNOT becomes a CZ gate between Hadamard-type single-qubit rotations. The compiled circuit has 9 CZ gates, 38 $\sqrt{X}$ gates and 50 $R_z$ gates, and a depth of 38. The parameters enter only through the $R_z$ angles. Hence the circuit was compiled once, and each SPSA iteration only bound new parameter values. Gate twirling and dynamical decoupling were added by the Runtime Sampler at execution and are not shown.

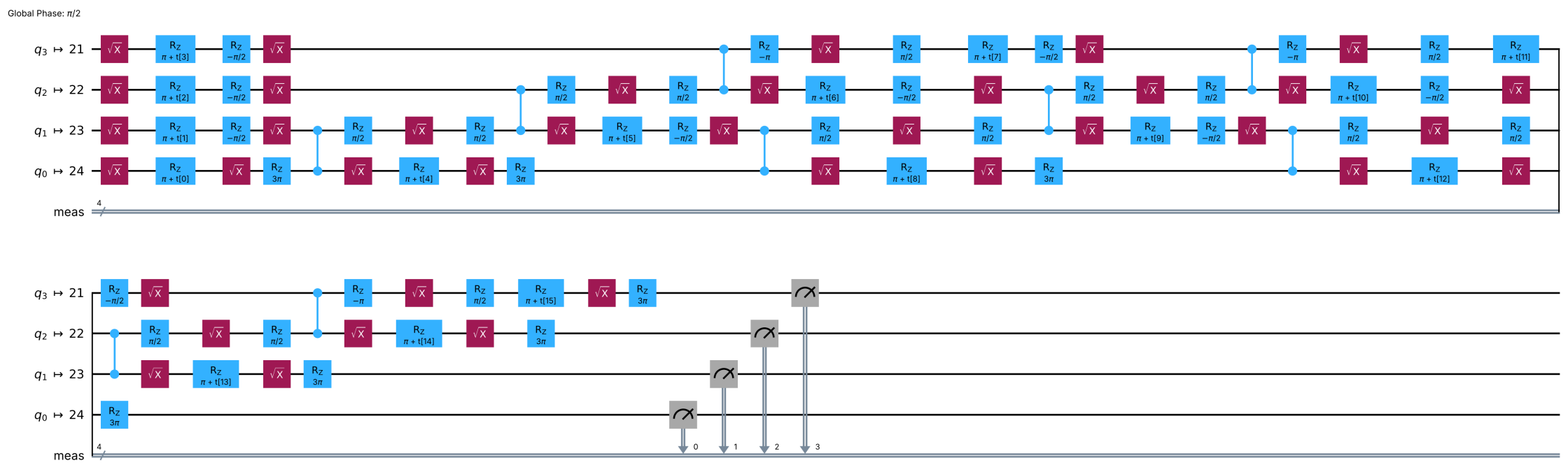

***Figure F2.*** *The circuit of Fig. F1 transpiled for ibm_cleveland on the physical qubits (24, 23, 22, 21). The labels* $q_i \mapsto k$ *give the physical qubit;* $t[k]$ *denotes* $\theta_k$*.*

Table F1 lists the parameters. The classically trained values reproduce the grid ground state with a fidelity of 1.0000. The initial values of the on-device VQE were obtained by adding Gaussian perturbations with a standard deviation of 0.3 rad (random seed 0), which gives a noiseless energy error of 4,939 bp. After 150 SPSA iterations the noiseless energy error is 130 bp. The final parameters do not return exactly to the trained values. The parameters $\theta_0$, $\theta_4$ and $\theta_8$ of the most significant qubit differ from them by 0.6–0.9 rad, whereas the final-layer parameters $\theta_{12}$–$\theta_{15}$ differ by less than 0.1 rad.

**Table F1.** Parameters (rad) of the four-qubit Vasicek ansatz: classically trained values, initial values of the on-device VQE, and values after 150 SPSA iterations on ibm_cleveland.

| Parameter | Layer | Qubit | Trained | Initial | After on-device VQE |
|---|---|---|---|---|---|
| $\theta_0$ | 1 | $q_0$ | -3.142 | -3.104 | -4.007 |
| $\theta_1$ | 1 | $q_1$ | +4.192 | +4.152 | +3.779 |
| $\theta_2$ | 1 | $q_2$ | -1.949 | -1.757 | -1.825 |
| $\theta_3$ | 1 | $q_3$ | +1.595 | +1.626 | +1.663 |
| $\theta_4$ | 2 | $q_0$ | +0.000 | -0.161 | -0.600 |
| $\theta_5$ | 2 | $q_1$ | -1.236 | -1.127 | -0.942 |
| $\theta_6$ | 2 | $q_2$ | +1.110 | +1.502 | +1.116 |
| $\theta_7$ | 2 | $q_3$ | +0.547 | +0.831 | +0.955 |
| $\theta_8$ | 3 | $q_0$ | -4.712 | -4.924 | -5.356 |
| $\theta_9$ | 3 | $q_1$ | +0.388 | +0.008 | +0.252 |
| $\theta_{10}$ | 3 | $q_2$ | +2.274 | +2.087 | +2.305 |
| $\theta_{11}$ | 3 | $q_3$ | +0.886 | +0.899 | +1.137 |
| $\theta_{12}$ | final | $q_0$ | +3.142 | +2.444 | +3.193 |
| $\theta_{13}$ | final | $q_1$ | -3.142 | -3.207 | -3.186 |
| $\theta_{14}$ | final | $q_2$ | -3.142 | -3.515 | -3.218 |
| $\theta_{15}$ | final | $q_3$ | -0.000 | -0.220 | -0.083 |